\documentclass[twocolumn, prl,preprintnumbers]{revtex4-1}
\usepackage{amsmath,amssymb,graphicx,booktabs,bm,psfrag,color,slashed}

\begin{document}

\title{\boldmath Axion-like particles, lepton-flavor violation\\
and a new explanation of $a_\mu$ and $a_e$}

\preprint{CERN-TH-2019-124, IPPP/19/64, MITP/19-053}
\preprint{\today}

\author{Martin Bauer$^a$}
\author{Matthias Neubert$^{b,c}$}
\author{Sophie Renner$^{b}$}
\author{Marvin Schnubel$^{b}$}
\author{Andrea Thamm$^d$}

\affiliation{${}^a$Institute for Particle Physics Phenomenology, Department of Physics, Durham University, Durham,
DH1 3LE, UK\\
${}^b$PRISMA$^+$\! Cluster of Excellence, Johannes Gutenberg University, 55099 Mainz, Germany\\
${}^c$Department of Physics {\em\&} LEPP, Cornell University, Ithaca, NY 14853, U.S.A.\\
${}^d$Theoretical Physics Department, CERN, 1211 Geneva, Switzerland}

\begin{abstract}
Axion-like particles (ALPs) with lepton-flavor violating couplings can be probed in exotic muon and tau decays. The sensitivity of different experiments depends strongly on the ALP mass and its couplings to leptons and photons. For ALPs that can be resonantly produced, the sensitivity of three-body decays such as $\mu\to 3e$ and $\tau\to 3\mu$ exceeds by many orders of magnitude that of radiative decays like $\mu\to e\gamma$ and $\tau\to\mu\gamma$. Searches for these two types of processes are therefore highly complementary. We discuss experimental constraints on ALPs with a single dominant lepton-flavor violating coupling. Allowing for one or more such couplings offers qualitatively new ways to explain the anomalies related to the magnetic moments of the muon and electron.
\end{abstract}

\maketitle

{\em I. Introduction.} 
Axion-like particles (ALPs) can be the low-energy remnants of an ultraviolet (UV) extension of the Standard Model (SM) with a spontaneously broken approximate global symmetry \cite{Kim:1986ax}. Being pseudo Nambu-Goldstone bosons, the couplings of ALPs to SM particles are determined by the symmetry structure of the UV theory. A discovery could thus provide important information about new physics that is otherwise out of reach of collider experiments \cite{Jaeckel:2015jla,Brivio:2017ije,Bellazzini:2017neg,Bauer:2017ris,Bauer:2018uxu}. There is no strong theoretical reason for a given SM extension to respect the SM flavor structure. Indeed, the UV theory could even be responsible for the breaking of the SM flavor symmetries, in which case the ALPs are known as \emph{flavons} (or \emph{familons}) \cite{Wilczek:1983wf,Anselm:1985bp,Feng:1997tn,Bauer:2016rxs}. If the flavon has a coupling to gluons it could also explain the strong CP problem \cite{Calibbi:2016hwq,Ema:2016ops}. Rare flavor-violating meson decays are some of the most powerful probes of these models \cite{Bjorkeroth:2018dzu,Gavela:2019wzg,Bauer:2019long}. Besides potential tree-level flavor-violating ALP couplings, even flavor-conserving couplings are strongly constrained through ALP mixing with pseudoscalar mesons and loop-induced ALP couplings, which inherit the SM flavor structure \cite{Gavela:2019wzg,Bauer:2019long,Izaguirre:2016dfi,Freytsis:2009ct}. In the SM, lepton flavor-changing decays are suppressed by the neutrino mass-squared differences and predictions for $\text{Br}_\text{th}(\mu\to 3e)\sim\text{Br}_\text{th}(\mu\to e\gamma)\sim 10^{-55}$ \cite{Petcov:1976ff,Hernandez-Tome:2018fbq} are many orders of magnitude smaller than the experimental limits $\text{Br}_\text{exp}(\mu\to 3e)<1.0\cdot10^{-12}$ \cite{Bellgardt:1987du} and $\text{Br}_\text{exp}(\mu\to e\gamma)<4.2\cdot10^{-13}$ \cite{TheMEG:2016wtm}. The future experiments MEG\,II and Mu3e will increase the sensitivity by up to four orders of magnitude, reaching unprecedented precision in searching for new physics \cite{Baldini:2013ke,Blondel:2013ia}. 

In effective theories, the decays $\mu\to e\gamma$ and $\mu\to 3e$ can be induced by dipole and 4-fermion operators,
\begin{align}
   \mathcal{L} = \frac{C_1}{\Lambda^2}\,m_\mu \bar\mu\sigma_{\mu\nu} F^{\mu\nu} e
    + \frac{C_2}{\Lambda^2}\,(\bar\mu\,\Gamma_1 e)(\bar e\,\Gamma_2 e) \,.
\end{align}
The resulting $\mu\to 3e$ rate is strongly suppressed with respect to the $\mu\to e\gamma$ rate, unless the coefficient $C_2$ is large enough to overcome the phase-space suppression of the three-body decay \cite{Calibbi:2017uvl,deGouva:2009zz}. For example, for $|C_1|\gg|C_2|$ one finds $\text{Br}(\mu\to 3e)\sim 5\cdot 10^{-3}\,\text{Br}(\mu\to e\gamma)$. 
Searches for $\mu \to e\gamma$ therefore seem to provide a universal tool to find new physics in this sector. 

In this Letter we show how this expectation can break down for light new physics. Light ALPs can be produced resonantly in the two-body decay $\mu\to e a$. Thus, searches for $\mu\to 3e$ provide the most sensitive probe for ALPs in the mass range $2m_e<m_a<m_\mu-m_e$, if ALPs predominantly decay into $e^+e^-$ pairs. If ALPs decay into photons, the resonantly enhanced decay $\mu\to e a$ with $a\to\gamma\gamma$ also leads to a strong limit in this mass range. For very collinear photons, the finite detector resolution can result in a $\mu\to e\gamma_\text{eff}$ signal, where $\gamma_\text{eff}$ refers to a photon pair reconstructed as a single photon. The rate for this process dominates over the ALP-induced $\mu\to e\gamma$ rate by many orders of magnitude. Therefore, constraints from $\mu\to e\gamma$ are only relevant for $m_a>m_\mu$. In our analysis we compute the $\mu \to e\gamma^*$ form factors at arbitrary $q^2$. For the process $\mu\to 3e$ we include these contributions together with the tree-level ALP exchange. In an analogous way, we further discuss the sensitivity of searches for flavor-changing $\tau$-lepton decays induced by ALPs.

ALPs with flavor-conserving couplings to leptons have been proposed as a possible explanation for the $3.7\sigma$ deviation between the SM prediction and measurements of the anomalous magnetic moment of the muon, at the expense of introducing a very large ALP-photon coupling \cite{Chang:2000ii,Marciano:2016yhf,Bauer:2017ris}. Here we show that addressing also the recently reported $2.4\sigma$ tension in the anomalous magnetic moment of the electron in the same model requires ALP couplings to electrons and muons of very different magnitude and opposite sign. We then explore new and qualitatively different solutions to both the $a_\mu$ and $a_e$ anomalies by allowing for flavor off-diagonal ALP-lepton couplings. 

\begin{figure}
\begin{center}
\includegraphics[width=0.42\textwidth]{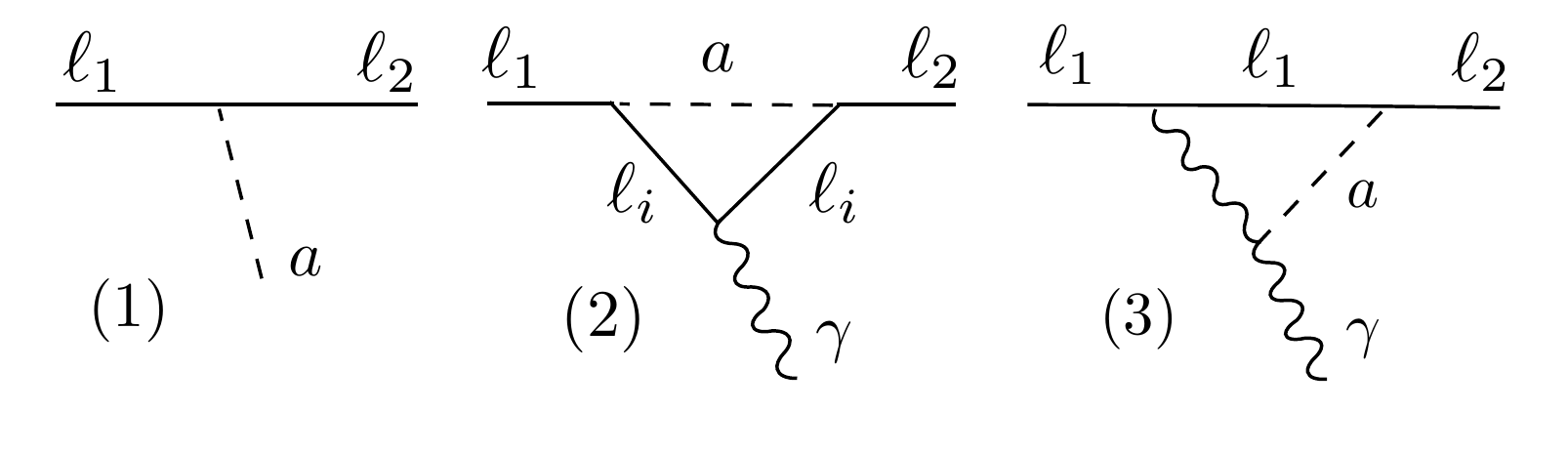}
\end{center}
\vspace{-0.5cm}
\caption{\label{fig:diagrams} 
Representative Feynman diagrams for ALP-induced $\ell_1\to\ell_2 a$ and $\ell_1\to\ell_2\gamma$ transitions.}
\end{figure}

\vspace{2mm}
{\em II. ALPs with lepton-flavor violating couplings.} 
General couplings of an ALP to charged leptons and photons are described by the effective Lagrangian
\begin{align}
   {\cal L}_{\rm eff} 
   = \frac{\partial^\mu a}{f} \Big( \bar\ell_L\bm{k}_E\gamma_\mu\ell_L + \bar\ell_R\bm{k}_e\gamma_\mu\ell_R \Big)
    + c_{\gamma\gamma} \frac{\alpha\,}{4\pi}\,\frac{a}{f}\,F_{\mu\nu}\tilde F^{\mu\nu} ,
\end{align}
where $\ell=(e,\mu,\tau)^T$, $f$ is the ALP decay constant, and we define the hermitian matrices $\bm{k}_E$ and $\bm{k}_e$ in the mass basis. The flavor-diagonal ALP couplings are given by the combinations $c_{\ell_i\ell_i}=(k_e)_{ii}-(k_E)_{ii}$. For the flavor off-diagonal couplings we define
\begin{align}\label{eq:cmue}
   c_{\ell_i\ell_j} = \sqrt{\left|(k_E)_{ij}\right|^2 + \left|(k_e)_{ij}\right|^2} \,.
\end{align}
Even if $c_{\gamma\gamma}=0$, ALP couplings to photons are induced at one-loop order \cite{Bauer:2017ris} and give rise to
\begin{align}\label{eq:1loopphoton}
   c_{\gamma\gamma}^\text{eff} = c_{\gamma\gamma} + \sum_i c_{\ell_i\ell_i} B_1(\tau_{\ell_i}) \,,
\end{align}
where $\tau_{\ell_i}=4m_{\ell_i}^2/m_a^2-i\epsilon$. The loop function is well approximated by $B_1(\tau)\approx 1$ for $\tau\ll1$ and $B_1(\tau)\approx -1/(3\tau)$ for $\tau\gg 1$, implying that effectively $c_{\gamma\gamma}^\text{eff}$ receives a contribution $c_{\ell_i\ell_i}$ from each lepton lighter than the ALP. Additional contributions are induced if the ALP couples to gluons or quarks.

If an ALP with lepton-flavor changing couplings to muons and electrons is light enough to be produced in muon decay, it can mediate the resonant decays $\mu\to e a\to 3e$  and $\mu\to e a\to e\gamma\gamma$ via diagram (1) in Fig.~\ref{fig:diagrams}. In the narrow-width approximation and dropping terms of order $m_e^2/m_\mu^2$, the corresponding decay rates are 
\begin{align}
   \Gamma(\mu\to e X)
   = \frac{m_\mu^3}{ 32\pi f^2}\,c_{e\mu}^2\,\bigg(1-\frac{m_a^2}{m_\mu^2}\bigg)^2\,\text{Br}(a\to X) \,,
\end{align}
where $X=e^+e^-,\gamma\gamma$, and the relevant ALP branching fractions can be computed using the partial decay widths of the ALP into electrons and photons, given by
\begin{align}\label{eq:alpdecay}
   \Gamma(a\to e^+e^-) 
   &= \frac{m_a m_e^2}{8\pi f^2}\,|c_{ee}|^2 \sqrt{1-\frac{4m_e^2}{m_a^2}} \,, \notag\\
   \Gamma(a\to \gamma\gamma) 
   &= \frac{\alpha^2 m_a^3}{64\pi^3f^2}\,\big|c_{\gamma\gamma}^\text{eff}\big|^2 .
\end{align}
If only ALP couplings to leptons appear in the UV theory and the ALP-photon coupling is induced through \eqref{eq:1loopphoton}, then the decay into photons is suppressed, $\text{Br}(a\to\gamma\gamma)\approx\alpha^2m_a^2/(8\pi^2 m_e^2)\,\text{Br}(a\to e^+e^-)$ for $m_e\ll m_a\ll m_\mu$. 

For $m_a>m_\mu$ we compute the $\mu\to 3e$ decay rate taking into account both the $\mu\to e a^*\to 3e$ and $\mu\to e\gamma^*\to 3e$ subprocesses and their interference. Since the ALP in the first subprocess is now off-shell, the corresponding amplitude is suppressed by the electron mass and is no longer dominant. The $\mu\to e\gamma^*$ amplitude can be described in terms of six $q^2$-dependent form factors, which we have computed analytically from diagrams (2) and (3) in Fig.~\ref{fig:diagrams}. Explicit expressions will be given elsewhere \cite{Bauer:2019long}. Two of these form factors evaluated at $q^2=0$ determine the $\mu\to e\gamma$ decay rate, for which we obtain (neglecting terms suppressed by $m_e^2/m_\mu^2$)
\begin{align}
   \Gamma(\mu\to e\gamma) 
   = \frac{\alpha m_\mu^5\,c_{e\mu}^2}{4096\pi^4 f^4}
    \left| c_{\mu\mu}\,g_1(x) + \frac{\alpha}{\pi}\,c_{\gamma\gamma}^\text{eff}\,g_2(x) \right|^2\! ,
\end{align}
where $x=m_a^2/m_\mu^2-i\epsilon$. The loop functions read
\begin{align}
   g_1(x) &= 2\sqrt{4-x}\,x^{\frac32} \arccos\frac{\sqrt{x}}{2} + 1 - 2x
    + \frac{3-x}{1-x}\,x^2\ln x \,,\notag\\[-5pt]
   g_2(x) &= 2\ln\frac{\Lambda^2}{m_\mu^2} - 2 - \frac{x^2\ln x}{x-1} + (x-1)\ln(x-1) \,,
\end{align}
where $\Lambda=4\pi f$ is the UV cutoff, and $g_1(x)$ agrees with the result of a double parameter integral derived in \cite{Lindner:2016bgg}. For simplicity, we have neglected the contributions with a $\tau$-lepton in the loop ($\ell_i=\tau$), which involve two flavor-changing parameters and are likely to be subdominant.

\begin{figure*}
\begin{center}
\includegraphics[width=0.93\textwidth]{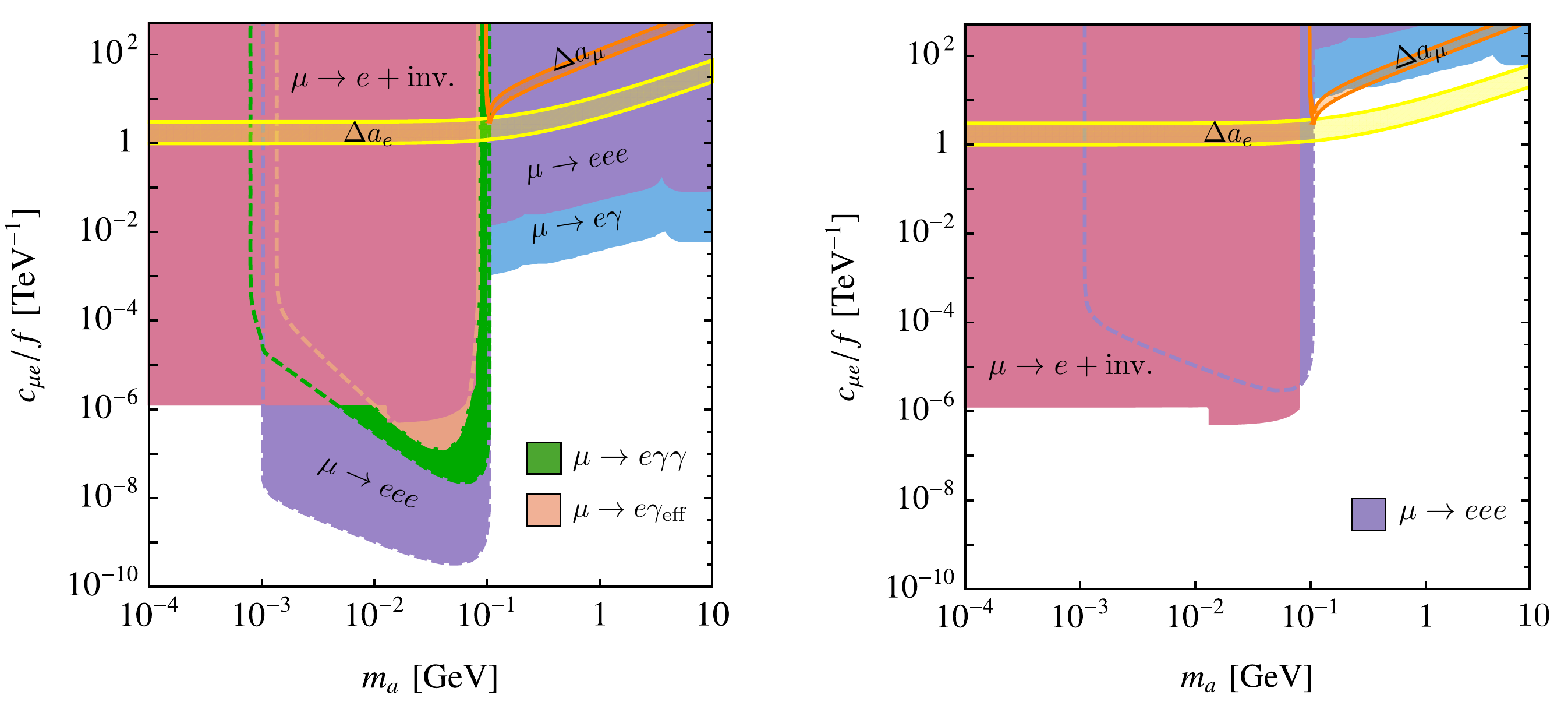}
\end{center}
\vspace{-0.5cm}
\caption{\label{fig:mue} 
Present experimental constraints on the effective ALP coupling to muons and electrons ($c_{e\mu}$) assuming universal couplings $c_{\ell\ell}/f=1/$TeV (left panel) and $c_{\ell\ell}/f=10^{-4}/$TeV (right panel). The parameter space for which $\Delta a_e$ and $\Delta a_\mu$ can be explained is shown in yellow and orange, respectively (see Sec.~IV).}
\end{figure*}

\begin{figure*}
\begin{center}
\includegraphics[width=0.93\textwidth]{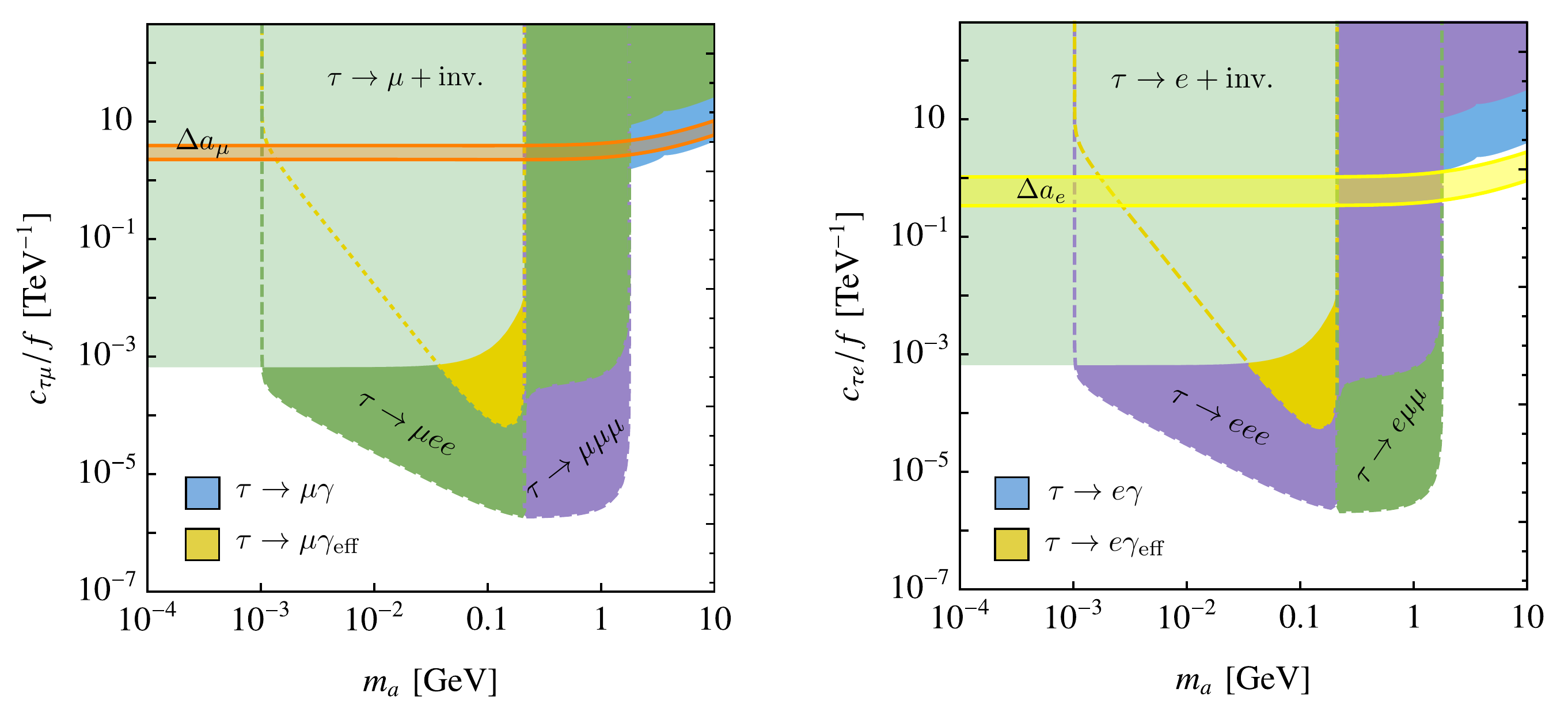}
\end{center}
\vspace{-0.5cm}
\caption{\label{fig:taus} 
Present experimental constraints on the effective ALP couplings to muons and taus (left panel) and electrons and taus (right panel), assuming $c_{\ell\ell}/f=1/$TeV. The parameter space for which $\Delta a_e$ and $\Delta a_\mu$ can be explained is shown in yellow and orange, respectively (see Sec.~IV).}
\end{figure*}

\vspace{2mm}
{\em III. Constraints on ALPs from lepton-flavor violation.} 
In the following we define conservative benchmark scenarios, in which the ALP-photon coupling in the UV theory vanishes ($c_{\gamma\gamma}=0$). We assume universal flavor-diagonal ALP couplings to leptons $c_{\ell_i\ell_i}\equiv c_{\ell\ell}$, a loop-induced coupling to photons and a single flavor-violating coupling $c_{\ell_1\ell_2}\ne 0$. In Fig.~\ref{fig:mue} we show the corresponding constraints on an ALP for the case of $c_{\ell\ell}/f=1/$TeV (left panel) and $c_{\ell\ell}/f=10^{-4}/$TeV (right panel), and a flavor-violating coupling $c_{e\mu}$ for a wide range of ALP masses. For very light ALPs ($m_a<2m_e$), the strongest constraint arises from a search for $\mu\to e$ decays with missing energy by TWIST \cite{Bayes:2014lxz}. ALP decays into photons are possible in this mass range, but according to \eqref{eq:alpdecay} the ALP decay width is strongly suppressed, leading to a long lifetime. For $2m_e<m_a<m_\mu$, the constraint derived from searches by SINDRUM for $\mu\to 3e$ is by far the strongest \cite{Bellgardt:1987du}, because this decay is enhanced from the ALP going on-shell in this mass range. The analogous bounds derived from a Crystal Ball search \cite{Bolton:1988af} for the decay $\mu\to e\gamma\gamma$ are less stringent, because in our scenario the ALP coupling to photons is loop-induced. The sensitivity from searches for $\mu\to e\gamma$ is enhanced in this parameter space as well, if the photons in $\mu\to e\gamma\gamma$ are collinear and cannot be distinguished from a single photon $\gamma_\text{eff}$ in the detector. In deriving these bounds we have taken into account the macroscopic ALP decay length, which implies that only a fraction of all decays can be reconstructed in the detector \cite{Bauer:2017ris}. Together with the $m_a$-dependence of the ALP lifetime governed by \eqref{eq:alpdecay} this explains the slopes of the relevant contours in Fig.~\ref{fig:mue} (see \cite{Bauer:2019long} for more details). Note also that in the presence of a tree-level coupling $c_{\gamma\gamma}\neq 0$ constraints from $a\to\gamma\gamma$ decays would be strengthened, whereas the bounds derived from $\mu\to 3e$ decay would get weaker.

For ALP masses $m_a>m_\mu$, the most important bound follows from the search for $\mu\to e\gamma$ by MEG \cite{TheMEG:2016wtm}. The right panel of Fig.~\ref{fig:mue} shows the corresponding bounds for a much smaller value of the flavor-diagonal lepton coupling $c_{\ell\ell}/f=10^{-4}/$TeV. While the $\mu\to e+\text{invisible}$ constraint remains largely unaffected, the remaining constraints get relaxed by about a factor of $10^4$ compared with the left panel. In the intermediate mass range $2m_e<m_a<m_\mu$, the reason is that the fraction of events reconstructed in the detector scales (approximately) with $\tau_a^{-1}\propto (c_{\ell\ell}/f)^2$ \cite{Bauer:2017ris}. For heavier masses $m_a>m_\mu$ the ALP lifetime is irrelevant, but the $\mu\to e\gamma$ and $\mu\to 3e$ decay rates scale with $(c_{\ell\ell}/f)^2$.

The above discussion shows that various searches for lepton-flavor violating ALP couplings are highly complementary and cover different regions in the parameter space spanned by the ALP mass and its couplings to leptons and photons. Future searches for $\mu\to e\gamma$ \cite{Baldini:2013ke} and $\mu\to 3e$ \cite{Blondel:2013ia} will allow one to strengthen the derived bounds significantly. 

In Fig.~\ref{fig:taus} we repeat the above analysis for lepton-flavor violating ALP couplings $c_{\tau\mu}$ (left panel) and $c_{\tau e}$ (right panel), this time considering universal couplings $c_{\ell\ell}/f=1/$TeV only. The plots show a similar structure as in the left panel of Fig.~\ref{fig:mue}. The strongest bounds for $m_a<2m_e$ are obtained from searches for the decays $\tau\to\mu+\text{invisible}$ and $\tau\to e+\text{invisible}$ by ARGUS \cite{Albrecht:1995ht}. ALPs with masses in the range $2m_e<m_a<m_\tau$ decay resonantly into lepton pairs, and the strongest constraints follow from searches for the three-body decays $\tau\to\mu ee$ and $\tau\to 3\mu$, or $\tau\to 3e$ and $\tau\to e\mu\mu$, performed by Belle \cite{Hayasaka:2010np}. For larger masses $m_a>m_\tau$, BaBar searches for the radiative decays $\tau\to\mu\gamma$ and $\tau\to e\gamma$ \cite{Aubert:2009ag} provide the only relevant constraints. 

\vspace{2mm}
{\em IV. ALP explanations for $\Delta a_\mu$ and $\Delta a_e$.} 
The SM prediction for the anomalous magnetic moment of the muon, $a_\mu=(g-2)_\mu/2$, deviates from the current best measured value by $3.7\sigma$ \cite{Bennett:2006fi,Keshavarzi:2018mgv}. The electron anomalous magnetic moment $a_e$ shows a tension of $2.4\sigma$ \cite{Hanneke:2008tm,Hanneke:2010au} after taking into account the recent, improved measurement of the fine-structure constant \cite{Parker:2018vye}. Interestingly, these deviations have opposite signs: $\Delta a_\mu=a_\mu^{\rm exp}-a_\mu^{\rm SM}=(27.06\pm 7.26)\cdot 10^{-10}$ and $\Delta a_e=a_e^{\rm exp}-a_e^{\rm SM}=(-87\pm 36)\cdot 10^{-14}$. 

The ALP-induced contribution from diagram (2) in Fig.~\ref{fig:diagrams} with flavor-diagonal couplings ($\ell_1=\ell_i=\ell_2=\mu$) has the wrong sign to explain $\Delta a_\mu$, whereas the contribution from diagram (3) can have either sign. Including both terms, one finds \cite{Chang:2000ii,Marciano:2016yhf,Bauer:2017ris}
\begin{align}\label{eq:amudiag}
   \Delta a_\mu = - \frac{m_\mu^2 c_{\mu\mu}^2}{16\pi^2f^2}
    \bigg[ h_1(x) + \frac{2\alpha}{\pi}\,\frac{c_{\gamma\gamma}^\text{eff}}{c_{\mu\mu}} 
    \bigg( \ln\frac{\Lambda^2}{m_\mu^2} - h_2(x) \bigg) \bigg] .
\end{align}
The loop functions are positive and satisfy $h_{1,2}(0)=1$ as well as $h_1(x)\approx (2/x)(\ln x-\frac{11}{6})$ and $h_2(x)\approx (\ln x+\frac{3}{2})$ for $x =m_a^2/m_\mu^2\gg 1$ \cite{Bauer:2017ris}. For very large ALP couplings to photons, $-c_{\gamma\gamma}^\text{eff}/c_{\mu\mu}\sim 10\!-\!30$, the second term in \eqref{eq:amudiag} can overcome the first one and explain $\Delta a_\mu$. Here we point out that such a large coupling can be induced at one-loop order through \eqref{eq:1loopphoton}, assuming non-universal ALP-lepton couplings $-c_{ee}/c_{\mu\mu}\approx 10\!-\!30$ and $m_a>2m_e$. Incidentally, an ALP coupling to electrons of this magnitude can also explain $\Delta a_e$ via a formula analogous to \eqref{eq:amudiag}. For example, with $m_a=0.5$ GeV, $c_{ee}/f=95/$TeV and $c_{\mu\mu}/f=-10/$TeV, we obtain $\Delta a_\mu=27.1\cdot 10^{-10}$ and $\Delta a_e=-84.5\cdot 10^{-14}$, both in agreement with experiment.

\begin{figure}
\begin{center}
\includegraphics[width=0.4\textwidth]{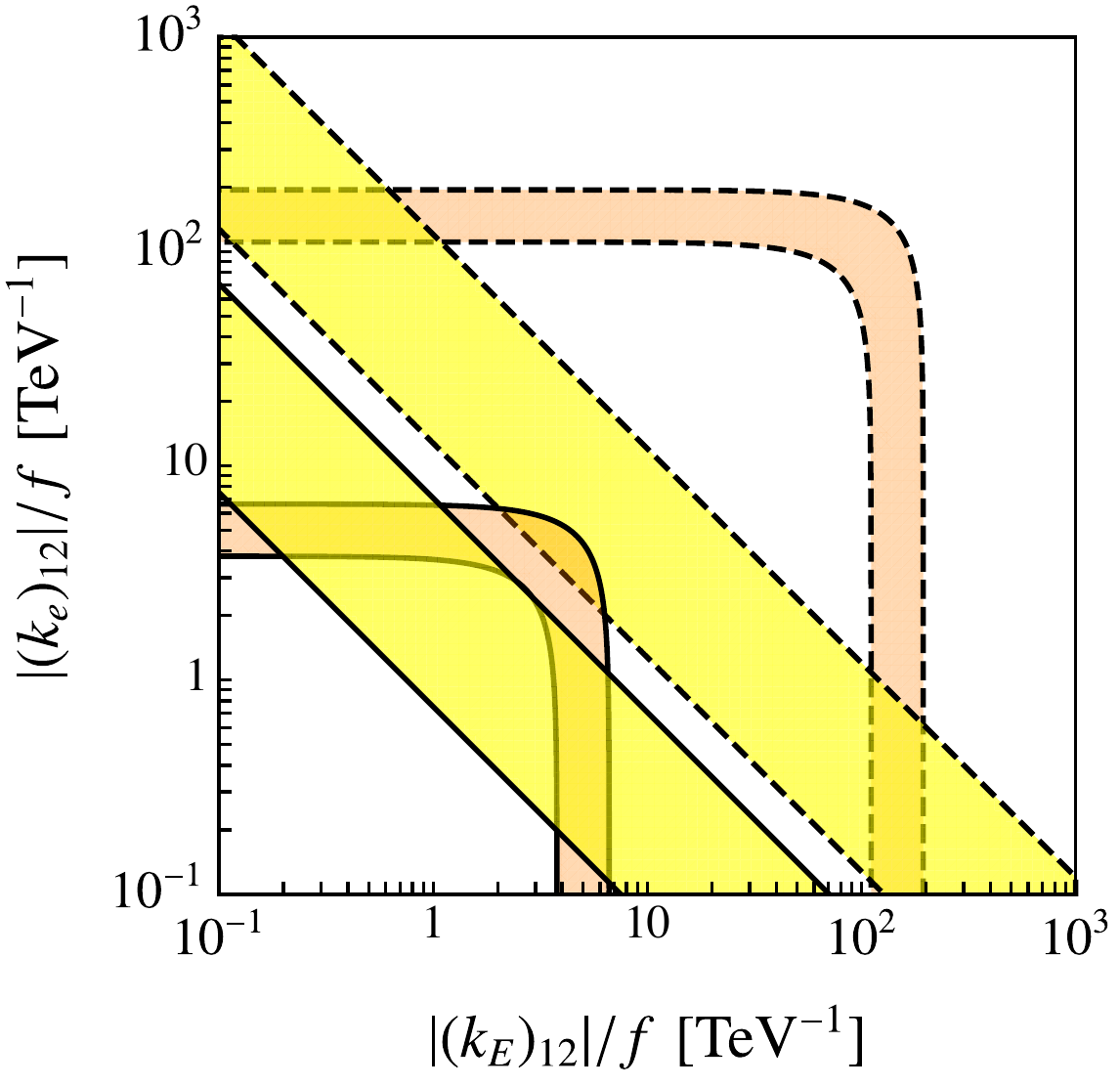}
\end{center}
\vspace{-0.5cm}
\caption{\label{fig:amue} 
Values of $(k_E)_{12}$ and $(k_e)_{12}$ for which $\Delta a_e$ (yellow) and $\Delta a_\mu$ (orange) can be explained for $m_a=110$ MeV (solid contours) and $m_a=1.5$ GeV (dashed contours). An explanation of $\Delta a_e$ requires $\text{Re}[(k_E)_{12} (k_e)_{12}^*]<0$.}
\end{figure}

As an intriguing alternative, we propose that dominant flavor-violating ALP couplings allow for a novel explanation of $\Delta a_\mu$ and $\Delta a_e$. The reason is that the contribution of the second diagram in Fig.~\ref{fig:diagrams} can have opposite sign depending on whether the lepton $\ell_i$ in the loop is lighter or heavier than the external lepton $\ell_1=\ell_2$. For the case of $\Delta a_\mu$ the diagram with the electron in the loop gives a positive contribution for $m_a>m_\mu$ given by (neglecting terms suppressed by $m_e^2/m_\mu^2$)
\begin{align}\label{eq:m2zero}
   \Delta a_\mu = \frac{m_\mu^2}{16\pi^2f^2}\,c_{e\mu}^2
    \bigg( x^2\ln\frac{x}{x-1} - x - \frac{1}{2} \bigg) .
\end{align}
On the other hand, the same couplings that enter the definition of $c_{e\mu}$ in \eqref{eq:cmue} lead to the contribution 
\begin{align}\label{eq:m1zero}
   \Delta a_e \!= \frac{m_em_\mu}{8\pi^2f^2}\,\text{Re}\big[(k_E)_{12} (k_e)_{12}^*\big]\!
    \bigg[ \frac{x^2\ln x}{(x-1)^3} - \frac{3x-1}{2(x-1)^2} \bigg] ,
\end{align}
which can be of either sign. In Fig.~\ref{fig:mue} we show the 95\% CL regions in which $\Delta a_\mu$ (orange) and $\Delta a_e$ (yellow) are explained in terms of these contributions. In deriving the corresponding bands we have assumed that $(k_E)_{12}=(k_e)_{12}=c_{e\mu}/\sqrt{2}$. While constraints sensitive to $c_{\mu\mu}$ are considerably weakened in the right panel, both \eqref{eq:m1zero} and \eqref{eq:m2zero} are independent of the flavor-diagonal coupling $c_{\ell\ell}$. As a result, for $m_a>m_\mu$ and (very) small flavor-diagonal ALP couplings to leptons either $\Delta a_\mu$ or $\Delta a_e$ can be explained by $c_{e\mu}/f>1/$TeV. A simultaneous explanation is possible if $(k_E)_{12}\ne(k_e)_{12}$, as illustrated in Fig.~\ref{fig:amue}. 

As a third possibility, either $\Delta a_\mu$ or $\Delta a_e$ can be explained by invoking flavor off-diagonal ALP couplings to $\tau$ leptons, which gives rise to contributions analogous to \eqref{eq:m1zero} with obvious substitutions. In Fig.~\ref{fig:taus}, we show the corresponding 95\% CL region in orange and yellow, respectively. This requires $m_a>m_\tau$ and, in the case of $\Delta a_\mu$, a flavor-diagonal ALP coupling $|c_{\tau\tau}|/f<0.3/$TeV. Note that a simultaneous explanation of both anomalies in terms of flavor-violating ALP couplings to $\tau$ leptons is not possible, because the contribution to the $\mu\to e\gamma$ decay arising from diagram (2) in Fig.~\ref{fig:diagrams} excludes this possibility. However, for small flavor-diagonal ALP couplings to leptons either $\Delta a_\mu$ or $\Delta a_e$ can be explained by $c_{e\mu}/f>1/$TeV (see above), and a contribution from $c_{\tau\mu}/f\sim 1/$TeV or $c_{\tau e}/f\sim 1/$TeV can explain the respective other anomaly. 

In this Letter we have shown that searches for lepton flavor-violating transitions provide highly complementary constraints on ALP couplings to leptons and photons. This strengthens the case for a broad program of experiments hunting for lepton flavor-violating decays. At the same time we have pointed out a possible connection between lepton flavor-violation and the observed tensions between theory and measurements of the muon and electron anomalous magnetic moments. We have discussed several ways in which ALPs with flavor non-universal couplings to leptons could explain these anomalies simultaneously.

\vspace{2mm}
{\em Acknowledgements.} 
We thank the Mainz Institute for Theoretical Physics (MITP) for hospitality, coffee and inspiring discussions  while conducting this research. This work has been supported by the Cluster of Excellence {\em Precision Physics, Fundamental Interactions, and Structure of Matter} (PRISMA$^+$\! EXC 2118/1) funded by the German Research Foundation (DFG) within the German Excellence Strategy (Project ID 39083149), and by grant 05H18UMKB1 of the German Federal Ministry for Education and Research (BMBF).

\end{document}